\documentclass{aa}
\usepackage{graphicx}
\def\aj{AJ}%
\def\apj{ApJ}%
\def\aap{A\&A}%
\def\mnras{MNRAS}%
\def\pasj{PASJ}%
\def\gx{GX~339$-$4}
\def\cx{Cyg~X$-$1}
\def\cb{Cyg~X$-$3}
\def\grs{GRS~1915$+$105}

\def\1e{1E~1740.7$-$2942}

\def\xteonze{XTE~J1118$+$480}
\def\vq{V404~Cyg}

\begin{document}
   \title{Radio / X-ray correlation in the low/hard state of GX~339--4 }

\titlerunning{Radio/X-ray flux correlation in \gx}


   \author{S. Corbel\inst{1}
          \and
          M.A. Nowak\inst{2}
          \and
          R.P. Fender\inst{3}
	  \and
	  A.K. Tzioumis\inst{4}
	 \and
	  S. Markoff\inst{5}
          }

   \offprints{S. Corbel}

   \institute{Universit\'e Paris VII and Service d'Astrophysique (F\'ed\'eration APC), 
		CEA Saclay, F-91191 Gif sur Yvette, France \\
		\email{corbel@discovery.saclay.cea.fr} 
	\and
		Chandra X-ray Science Center, Massachusetts Institute
		of Technology, NE80-6077, 77 Massachusetts Ave., 
                Cambridge MA 02139, USA\\
	\and
		Astronomical Institute `Anton Pannekoek', University
		of Amsterdam and Center for High Energy Astrophysics, 
                Kruislaan 403, 1098 SJ Amsterdam, The Netherlands \\
 	\and
                Australia Telescope National Facility, CSIRO, P.O. Box
                76, Epping NSW 1710, Australia\\
	\and
                Massachusetts Institute of Technology, 
		Center for Space Research, 
		NE80-6035, 77 Massachusetts Ave.,
                Cambridge MA 02139, USA\\
	}
		
   \date{Received: 6 November 2002; Accepted: 6 January 2003 }

   \abstract{We present the results of a long-term study of the black
hole candidate \gx\ using simultaneous radio (from the Australia
Telescope Compact Array) and X-ray (from the Rossi X-ray Timing
Explorer and BeppoSAX) observations performed between 1997 and 2000.
We find strong evidence for a correlation between these two emission
regimes that extends over more than three decades in X-ray flux, down
to the quiescence level of \gx. This is the strongest evidence to date
for such strong coupling between radio and X-ray emission.  We discuss
these results in light of a jet model that can explain the radio/X-ray
correlation. This could indicate that a significant fraction of the
X-ray flux that is observed in the low-hard state of black hole
candidates may be due to optically thin synchrotron emission from the
compact jet.

   }

    \maketitle
%

\section{Introduction}
 
Since its discovery in 1973 by the X-ray satellite {\em OSO-7}
(\cite{mar73}), the black hole candidate (BHC) \gx\ has been the
subject of many extensive studies from the radio bands to the hard
X-rays.  Nevertheless, the physical processes involved in its
broadband energy spectra have not been fully identified and
understood.  As the source has displayed a wide range of accretion
rates -- it is one of the rare sources that has been observed in all
canonical black hole X-ray states -- it is a prime target for studying
the accretion-ejection processes of accreting black hole systems
(persistent or transient). Among the canonical states, it is perhaps
the low/hard state (LHS) that has attracted the most attention in
recent years.  Observations of various components in the LHS spectral
energy distribution (SED) have highlighted analogies with the flat
spectra of low-luminosity active galactic nuclei (AGN) (\cite{fal96}).

Radio emission from \gx\ during the LHS is characteristic of a
self-absorbed compact jet (\cite{cor00}), similar to that considered
for flat spectrum AGNs (Blandford \& K$\ddot{\mathrm{o}}$nigl 1979).
Regular radio observations have shown that the compact jet of \gx\ was
quenched in the high/soft state (\cite{fen99,cor00}).  Similar
properties have now been observed in a growing number of persistent
and transient BHCs (the jet has even been resolved in \cx, Stirling et
al. 2001), thus suggesting that compact jets are ubiquitous in BHCs during the LHS
(\cite{fen01}).  In addition to being responsible for most of the
emission in the radio regime, the compact jets may have a significant
contribution in the infrared/optical bands (Corbel \& Fender 2002) and
could also be part of the processes involved in producing the X-ray
emission (\cite{mar01}). In that case, it would imply that compact
jets are very powerful and could dominate the entire SED of BHC during
their low/hard states.

The other main components of the SED are a possible thermal
contribution from the accretion disk, which extends from the
near-infrared/optical to the soft X-rays, and the possible
Comptonisation of accretion disk photons with a hot electron corona,
which likely contribute mostly in soft and hard X-rays
(\cite{now02,don02}). Despite not being observed, the companion star
in the \gx\ system is likely an evolved low-mass star
(\cite{sha01,cha02,cow02}).

A good way to assess the contribution of jets at high energy is to
perform a broadband study of these systems simultaneously at radio and
X-ray frequencies, and in particular to study the correlation that
could exist between these two emission domains.  
In fact, such a correlation has already been found for BHC in the LHS,
e.g., \gx\ (\cite{han98,cor00}) and
\cx\ (\cite{bro99}) for BHC in the LHS. These radio/X-ray
observations, however, only sampled a very limited range of X-ray and
radio fluxes (or accretion rates).  Most of these previous studies also
suffered from the lack of sensitivity of the X-ray observations.  A
similar correlation also has been observed in the hard state of \cb\
(\cite{mcc99,cho02}). A more complex relation, but still indicating a
relation between radio emitting electrons and the hard X-ray power law
dominated state (like that in the LHS of BHC), is found in \grs\
(\cite{kle02}).

An interpretation for such a tight correlation is that it is the
result of high energy synchrotron emission from the compact jet, as
already has been proposed for \xteonze\ (\cite{mar01}).
Alternatively, it possibly could be due to Compton scattering of
photons from the companion star or the accretion flow off of the jet's
leptons (\cite{geo02}).  In this paper, we report the results of a
long-term study of \gx\ performed simultaneously in radio and X-ray
during the years 1997-2000, in order to investigate the radio/X-ray
correlation over many orders of magnitude with sensitive observations.
Evidence for evolution of this correlation with energy is also
presented. We discuss these results in light of the jet model of
Markoff et al. (2003).

\begin{table*}[hbt!]
\centering
     $$
\begin{tabular}{lccccc}
            \hline
            \noalign{\smallskip}
         &  \multicolumn{4}{c}{X - ray flux } &   Radio flux density \\
            \noalign{\smallskip}
\cline{2-5}	
            \noalign{\smallskip}
\multicolumn{1}{c}{Date}	&  3 - 9 keV            & 9 - 20 keV            &  20 - 100 keV       & 100 - 200 keV           &   8.6 GHz      \\
            \noalign{\smallskip}
\multicolumn{1}{c}{(y.m.d.)}    &      \multicolumn{4}{c}{ (10$^{-10} \rm \, erg \,s^{-1} \,cm^{-2}$) } & (mJy) \\
            \noalign{\smallskip}
            \hline
            \noalign{\smallskip}
1997.02.03  &  10.74 $\pm$ 0.11               & 10.28 $\pm$ 0.10     & 29.05 $\pm$ 0.29   & 11.57   $\pm$ 0.12     & 9.10  $\pm$ 0.10 \\
1997.02.10  &  9.41  $\pm$ 0.09               & 9.10  $\pm$ 0.09     & 26.81 $\pm$ 0.27   & 10.74   $\pm$ 0.11     & 8.20  $\pm$ 0.10 \\
1997.02.17  &  9.02  $\pm$ 0.09               & 8.73  $\pm$ 0.09     & 25.50 $\pm$ 0.26   & 10.49   $\pm$ 0.11     & 8.70  $\pm$ 0.10 \\
1999.02.12  &  4.76  $\pm$ 0.05               & 4.21  $\pm$ 0.04     & 11.80 $\pm$ 0.12   & 4.16    $\pm$ 0.40     & 4.60  $\pm$ 0.08 \\
1999.03.03  &  4.75  $\pm$ 0.05               & 4.62  $\pm$ 0.05     & 14.91 $\pm$ 0.15   & 6.66    $\pm$ 0.32     & 5.74  $\pm$ 0.06 \\
1999.04.02  &  4.93  $\pm$ 0.05               & 4.90  $\pm$ 0.05     & 15.87 $\pm$ 0.16   & 8.75    $\pm$ 0.45     & 5.10  $\pm$ 0.07 \\
1999.04.22  &  2.30  $\pm$ 0.02               & 2.34  $\pm$ 0.02     & 7.37  $\pm$ 0.10   & 4.04    $\pm$ 0.36     & 3.11  $\pm$ 0.04 \\
1999.05.14  &  0.76  $\pm$ 0.01               & 0.74  $\pm$ 0.01     & 2.23  $\pm$ 0.17   & 1.39    $\pm$ 0.29     & 1.44  $\pm$ 0.04 \\
1999.06.25  &  0.059 $\pm$ 0.006              & 0.052 $\pm$ 0.005    & $<$  0.17          & $<$  0.29      & 0.24  $\pm$ 0.05 \\
1999.07.07  &  0.029 $\pm$ 0.002              &  $<$  0.01           & $<$  0.17          & $<$  0.33      & 0.12  $\pm$ 0.04 \\
1999.07.29  &  $<$  0.008                    &  $<$  0.012           & $<$  0.16          & $<$  0.29      & $<$  0.036 \\
1999.08.17$^{\mathrm{a}}$  &  0.025 $\pm$ 0.003   & 0.019 $\pm$ 0.014     & 0.110  $\pm$ 0.046   & $<$  0.05      & 0.27  $\pm$ 0.07 \\
1999.09.01$^{\mathrm{b}}$  &  0.037  $\pm$ 0.003  &  $<$  0.01            & $<$ 0.17            & $<$  0.29      & 0.32  $\pm$ 0.05 \\
2000.09.10$^{\mathrm{c,d}}$ &  0.0057 $\pm$ 0.001 &  $<$  0.0034          & $<$ 0.005           & $<$  0.07      & $<$  0.02 \\
            \noalign{\smallskip}
            \hline      
\end{tabular}
     $$
\begin{list}{}{}
\item[$^{\mathrm{a}}$] Flux (or upper limits) above 9 keV are deduced from the BeppoSAX observations performed on 1999.08.13
\item[$^{\mathrm{b}}$] Average X-ray flux, based on the PCA observations on 1999.08.28 and 1999.09.04
\item[$^{\mathrm{c}}$] Radio observations on 2000.09.12, 2000.09.15 and 2000.09.18
\item[$^{\mathrm{d}}$] X-ray observations performed by BeppoSAX
\end{list}
\caption[]{ Observing log of the simultaneous X-ray (PCA, unless otherwise noted) 
and radio (8.6 GHz) observations of \gx\ performed during a low/hard state.  X-ray absorbed 
fluxes are all normalized to PCA . Upper limits are given at the one sigma level.}
         \label{tab_res}
\end{table*}
 
\section{Observations}

\subsection{Radio observations}

All radio observations were performed with the Australia
Telescope Compact Array (ATCA). The ATCA synthesis telescope is an
east-west array consisting of six 22 m antennas. The 8.6 GHz data that
we used is from Corbel et al. (2000); however, we re-analaysed
all (five) observations for which the radio flux densities were weaker
than 1\,mJy. Further details concerning the ATCA and its data
reduction can be found in Corbel et al. (2000). We also added the
result of a series of three new ATCA observations, for a total
duration of nearly 20 hours, performed on 2000 September 12, 15 and
18, during the recent off state.  These observations provided a strong
(99\% confidence level) upper limit of 60 $\mu$Jy at 8.6 GHz, which is
the best constraint we have for the level of radio emission
originating from \gx\ during its off state.

\subsection{X-ray observations}
 
\subsubsection{RXTE}

We used the {\em Rossi X-ray Timing Explorer} ({\em RXTE}) to
perform a number of observations of \gx\ from 1997-1999, most of which
represent the LHS. Analysis of the brightest of these observations
was previously presented (Nowak et al. 2002); here we also
consider analyses of faint, ``off state'' observations from late 1999.
All data extractions were performed in an identical manner as that
described by Nowak et al. (2002).  The flux in various energy bands
was determined by fitting a model comprised of neutral hydrogen
absorption ($N_{\rm H}$ was fixed to $6\times10^{21}~{\rm cm^{-2}}$), a
multi-temperature disk blackbody (e.g., \cite{mitsuda84}) with peak
temperature fixed at 0.25\,keV, an exponentially cut-off broken power
law with break energy at $\approx 10$\,keV, and a (potentially) broad
Gaussian line with peak energy fixed at 6.4\,keV.  
The faintest observations were fit with a simpler absorbed, single power law, plus
(fixed peak energy) line feature.
Feng et al. (2001) found that the iron line is shifted to higher 
energies when \gx\ was observed at low X-ray fluxes. However, this shift
is not intrinsic to \gx, but is rather due to the Galactic diffuse emission
(\cite{wardz03}).
Note that due to
differences between the two sets of instruments that comprise {\em
RXTE}, the {\em Proportional Counter Array} ({\em PCA}, $\approx
3$-20\,keV) and the {\em High Energy X-ray Timing Experiment} ({\em
HEXTE}, $\approx 20$-200\,keV), a normalization constant between the
{\em PCA} and {\it HEXTE} detectors was used, and all flux values are
normalized to the {\em PCA} flux levels (for further descriptions of
this process, see Wilms et al. 1999). The flux error bars were chosen to be
the larger of the statistical error, or 1\%, which is a reasonable
estimate of the RXTE internal systematic error (e.g., Wilms et al.  1999). 
Short timescale ($\le$ few seconds) X-ray variability 
is usually observed in the low-hard state of \gx\ (\cite{smi99,now02}). However,
on a longer timescale (e.g. 10 minutes) the radio emission is steady (see Figure 3 in Corbel
et al. 2000) and so also is the X-ray spectrum of \gx\ integrated on those timescales (i.e.,
there is almost no very low frequency power in the power spectral densities, Nowak et al. 2002 and references therein).
Therefore the error bars used in Table 1
are likely not affected by the variability of the source (which is quite steady on 
timescales greater than 10 minutes).
Note that in Table~\ref{tab_res} we quote the {\em absorbed} flux level; however,
as we only consider energies $\ge 3$\,keV, this is at most a few
percent different to the unabsorbed flux level.

{\em RXTE} has a broad $\approx 1^\circ$ radius field of view, and
therefore is potentially subject to contamination from faint
background sources (or, in the case of \gx, diffuse emission from the
galactic ridge, \cite{wardz03}).  Four of the {\em RXTE} observations,
however, were performed simultaneously with the much narrower field of
view ($\approx 4$\,arcmin radius) {\em Advanced Satellite for
Cosmology and Astrophysics} ({\em ASCA}).  Utilizing the same models
described above, the 3-9\,keV flux of the brightest two simultaneous
observations determined by {\em ASCA} was 75-81\% of that
determined by the {\em PCA} -- consistent with a well-known
calibration offset between {\em PCA} and {\em ASCA} (see the
discussion in Nowak et al. 2002).  For the faintest two observations,
the relative normalizations of the {\em ASCA} spectra substantially
decreased with decreasing flux.  This was taken as evidence for a
faint background source or sources that lie within the field of view
of the {\em PCA}, but not within the field of view of {\em ASCA}.  The
{\em RXTE} observation of 1999 July 29 is assumed to be heavily
dominated by this contaminating source, and this spectrum, multiplied
by 0.78, is subtracted as a ``background correction'' before model
fitting and flux determination, from all {\em RXTE} observations
occurring later than the observation of May 14 1999.  With this
additional background subtracted, the {\em ASCA} determined fluxes of
the faintest two simultaneous observations become 73\% and 83\% of the
corrected {\em PCA} 3-9\,keV flux levels. Good agreement also is
obtained between the {\em ASCA} and the corrected {\em PCA} spectra.

\subsubsection{BeppoSAX}

During the recent off state of \gx\ we conducted an $\sim$ 50 ks
X-ray observation with {\em BeppoSAX} on September 10 2000.  \gx\ is
detected in the 1 - 10 keV energy range with both LECS and MECS, and
the spectrum can be fitted with a power-law with a photon index 2.22 $\pm$
0.24 (90 \% confidence level) with interstellar absorption fixed to 5.1 $\times\ 10^{21}$
cm$^{-2}$ ($\chi_0^2$ = 0.74 for 35 degrees of freedom). The absorbed
flux in the 3 - 9 keV energy range is 5.7 $\times\ 10^{-13}$ erg
cm$^{-2}$ s$^{-1}$ (relative to the MECS normalization). To within a few
percent, the fluxes normalized to MECS are consistent with the ones
normalized to PCA (e.g. Della Ceca et al. 2001). We
also re-analysed the BeppoSAX observation performed on
August 13 1999 by Kong et al. (2000), as it was close to the date of
one of our radio observations.  All measurements (radio and X-ray) are
tabulated in Table \ref{tab_res}.

\section{An extremely strong correlation between radio and X-ray
emissions during the low/hard state}

In Figures 1 to 4, we utilize the results from
Table~\ref{tab_res} and display the radio flux density at 8.6 GHz
versus the X-ray flux in different energy bands.  It is apparent from
these plots that a strong correlation exists between these two
emission regimes in \gx.  As all these measurements have been taken
during the low/hard state or during the transition to the off state
(which appears to be a weak luminosity version of the
low/hard state), we can further deduce that this correlation is a
property of the low/hard state.  We note that this strong
correlation extends over more than three orders of magnitude in X-ray
flux (e.g. the 3--9 keV band for which we have the best
coverage). The correlation appears to hold for the entire four year
period covered by the observations.  Specifically, the 1997
measurements lie on the same line as the measurements performed
during the transition to the off state in 1999 (Figs. 1 to 4), even though
there was a transition to the soft state between these sets of
observations (\cite{bell99,now02}).

In order to quantify the level of the correlation, we have calculated for
each of the X-ray bands the Spearman rank correlation coefficient (Barlow 1989)
between the radio and X-ray fluxes (Table \ref{tab_correl}). For this
calculation, all detections have been taken into account (i.e.
including even the points which are not strictly simultaneous or
affected by a small reflare in hard X-rays: 1999.08.17 and
1999.09.01).  It is clear from this analysis that the relationship
between the radio and the soft and hard X-ray fluxes is extremely strong
(a Student's t-test (Barlow 1989) shows the significance of the correlation is
greater than 99.9 \% for each of the energy bands).  This points to a
persistent coupling between the mechanism(s) of  the origin of the
radio and X-ray emissions while \gx\ is in the low/hard state.  This
is the strongest evidence to date for such a persistent
relation. 

   \begin{table}
     $$
         \begin{tabular}{cccc}
            \hline
            \noalign{\smallskip}
 X-ray band & r$_s$ & p & N \\
            \noalign{\smallskip}
            \hline
            \noalign{\smallskip}
3 - 9 keV     & 0.94 & 7.0 $\times\ 10^{-7}$ & 12 \\
9 - 20 keV    & 0.96 & 7.3 $\times\ 10^{-6}$ & 10 \\      
20 - 100 keV  & 0.97 & 2.2 $\times\ 10^{-5}$ & 9 \\      
100 - 200 keV & 0.95 & 2.6 $\times\ 10^{-4}$ & 8 \\      
            \noalign{\smallskip}
            \hline         
        \end{tabular}
     $$
      \caption[]{Spearman rank correlation coefficient, r$_s$, and the
two sided significance of its deviation from zero, p, between the radio
flux density measured at 8.6 GHz and the X-ray flux measured in
various energy bands. The number of data-points, N, used in the
calculations is also indicated in the last column.}
\label{tab_correl} \end{table}
 
It is possible to estimate a functional relationship between the
flux densities in radio (e.g. 8.6 GHz) and in the various X-ray
bands. A linear fit (on a log-log scale) is satisfactory for the four X-ray
bands. If we denote F$_{\mathrm{Rad}}$ as the radio flux density (in mJy)
at 8.64 GHz and F$_\mathrm{X}$ as the X-ray flux (in units of 10$^{-10}
\rm \, erg \,s^{-1} \,cm^{-2}$) in a given energy band, the relation
between these two fluxes can be expressed as $F_{\rm Rad} = a \times
F_{\rm X}^{b}$, where a and b are the two constant coefficients given in
Table \ref{tab_fit}.  These relations are valid while \gx\ is in the
standard low/hard state, as we recall that radio emission from \gx\ is
quenched in the high/soft state (\cite{fen99,cor00}).

As the same correlation appears to be maintained over this four year
period, it could be realistic to estimate the level of radio emission
from \gx\ by only measuring its X-ray flux. We note that little scatter is observed
around the fitting function for our measurements. This also indicates
that there is probably little time delay 
between the radio and X-ray emission. 
We also note the index $b$ apparently changes with X-ray energy band.
This is consistent with previous X-ray observations that indicate
that, with the hard state, the spectrum of \gx\ becomes spectrally
harder as the source becomes fainter (e.g., \cite{now02}).

   \begin{table}
     $$
         \begin{tabular}{ccc}
            \hline
            \noalign{\smallskip}
 X-ray band & a  & b \\
            \noalign{\smallskip}
            \hline
            \noalign{\smallskip}
3 - 9 keV     &  1.721 $\pm$ 0.035 &  0.706 $\pm$ 0.011 \\
9 - 20 keV    &  1.739 $\pm$ 0.035 &  0.715 $\pm$ 0.011 \\      
20 - 100 keV  &  0.667 $\pm$ 0.041 &  0.774 $\pm$ 0.021 \\      
100 - 200 keV &  1.024 $\pm$ 0.287 &  0.891 $\pm$ 0.104 \\      
            \noalign{\smallskip}
            \hline         
        \end{tabular}
     $$
     \caption[]{Parameters of the function used to fit the radio flux
density (in mJy) at 8.6 GHz, F$_{\mathrm{Rad}}$, versus the flux,
F$_\mathrm{X}$, measured in a given energy band ((in unit of
10$^{-10}\rm \, erg \,s^{-1} \,cm^{-2}$).  The relation is expressed
as $F_{\rm rad}= a~\times~F_{\rm X}^{b}$.}  \label{tab_fit}
\end{table}

\section{Discussions}

With this study, we showed that the previously observed correlation
between X-ray and radio fluxes during the low-hard state of \gx\
(\cite{han98,cor00}) extends over more than three decades in X-ray
luminosity. The observations, spread over almost four years, indicated
that the same fitting functions probably hold during these years,
despite an intervening state transition.  The radio emission in \gx\
is likely associated with the optically thick synchrotron emission
from the compact jet (\cite{cor00}).  The very strong correlation of
radio emission with X-ray flux indicates that synchrotron
processes may also play a role at high energies. 

It has already been pointed out for \gx\ that the near
infrared/optical bands show a spectral break that may indicate that
the X-ray spectrum is an extension of the optically thin synchrotron
emission from the compact jet (\cite{cor02}). Specifically, the infrared
points are consistent with an extrapolation of the slope from the
radio, while the infrared/optical break extrapolates to the observed
X-rays.  The jet model, originally developed for AGN and previously
applied to \xteonze\ (\cite{mar01}), has been further improved and
also applied to these datasets (with additional optical data), and is
discussed in a companion paper (\cite{mar03}). Markoff et al. (2003)
showed that the jet model can account for the broadband spectra of
\gx\, radio through optical through X-ray, primarily by only changing
two parameters: the jet power and the location of the acceleration
zone.  The fact that the correlation holds at very low X-ray
luminosity indicates that a compact jet is also produced when the
source is close to quiescence. Therefore, the broadband emission of
the jets has to be taken into account when studying the bolometric
luminosity of black hole in quiescence, e.g. Campana \& Stella (2000)
and Garcia et al. (2001).

It is interesting to note that the radio flux at 8.6 GHz is
proportional to the X-ray flux as $F_{\rm rad} \propto~F_{\rm
X}^{+0.71}$ (in the 3--9 and 9--20 keV bands) and that the same
behaviour with the same index $b$ has recently been found to hold for
the black hole transient \vq\ (\cite{gal02}).  The jet model of
Markoff et al. (2003) successfully explains this dependency
analytically using the formalism developed in Falcke \& Biermann
(1996). Indeed, if the only varying parameter of the model is the
power in the jet, then the X-ray flux is expected to vary as $F_{\rm
X} \propto~F_{\rm rad}^{+1.41}$, i.e. $F_{\rm rad} \propto~F_{\rm
X}^{+0.71}$, completely consistent with the behaviour of \gx\ and
\vq. The current versions of the jet model, however, do not 
account for the evolution of the exponent, $b$, in $F_{\rm rad}
\propto~F_{\rm X}^{b}$, for the higher energy X-ray bands, as is found
for \gx\ (Table \ref{tab_fit}). If a jet is the underlying cause of 
the overall radio/X-ray correlation, the model
may need to be further developed in order to explain this detailed
behavior.  However, it is important to point out that these higher
energies include the canonical "100 keV cutoff" region in the data,
where one expects the X-rays to decrease in comparison to the radio
emission.  However, the simplified accretion disk model used by
Markoff et al. (2003)  currently does not include all spectral components, e.g.,
the contribution from reflection which are often observed in the hard
state X-ray spectra of BHC (e.g., Done 2002).  Studies like these
clearly show, however, that including jet emission processes in any
models of BHC spectra in their low/hard state is crucially important
to fully understand the physical processes in these sources.

\clearpage

\begin{acknowledgements}
The Australia Telescope is funded by the Commonweath of Australia for
operation as a National Facility managed by CSIRO.  We would like to
thank Ben Chan and Richard Dodson for conducting the September 2000
ATCA observations.

\end{acknowledgements}

\clearpage
\newpage

 \begin{figure*}[hT!] \centering \includegraphics[width=14cm]{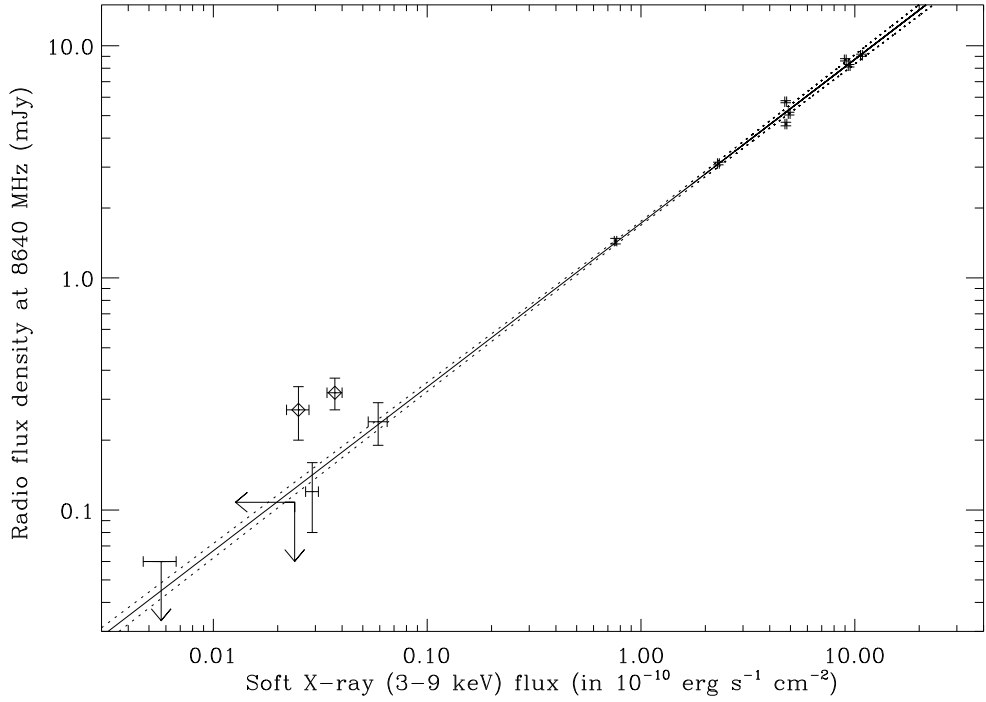}
	\caption{The radio flux density at 8.6 GHz is plotted versus the
   X-ray flux in the 3-9 keV energy band.
   The continuous line denotes the fit
   to the data with the function described in the body of the paper
   and with the parameters estimated in Table 3, the dotted line
   represents the one-sigma deviation to those parameters. Upper
   limits are plotted at the three sigma level. The diamond points are
   those points that are not strictly simultaneous (1999.08.17) or maybe
   affected by a small reflare observed in hard X-rays (1999.09.01, see
   Figure 15 in Corbel et al. 2000).  } \label{fig_fit} \end{figure*}

\begin{figure*}[hB!] \centering \includegraphics[width=14cm]{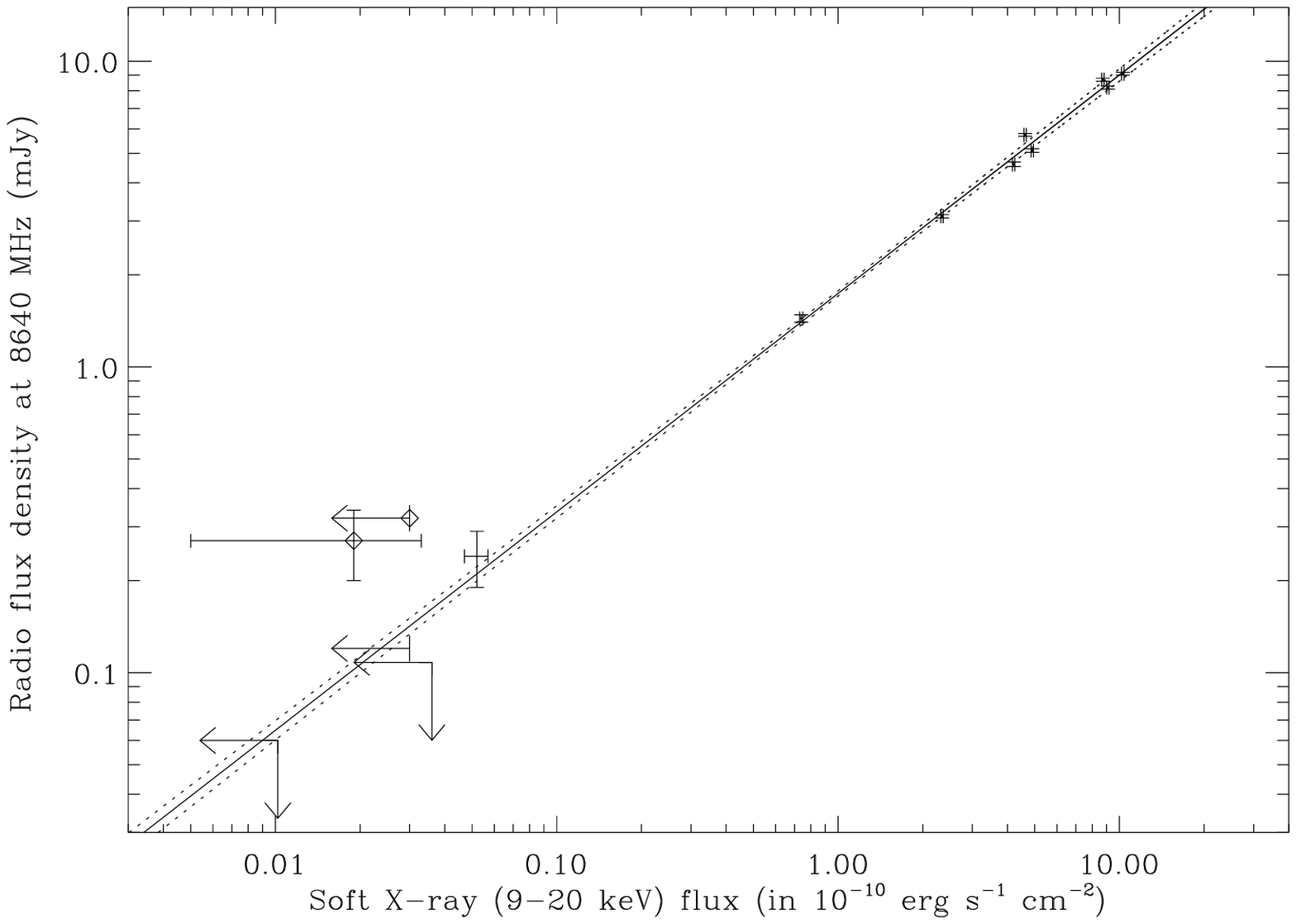}
\caption{Same as Fig. 1, but for the X-ray flux in the 9-20 keV energy band.}
\end{figure*}

\begin{figure*}[ht] \centering \includegraphics[width=14cm]{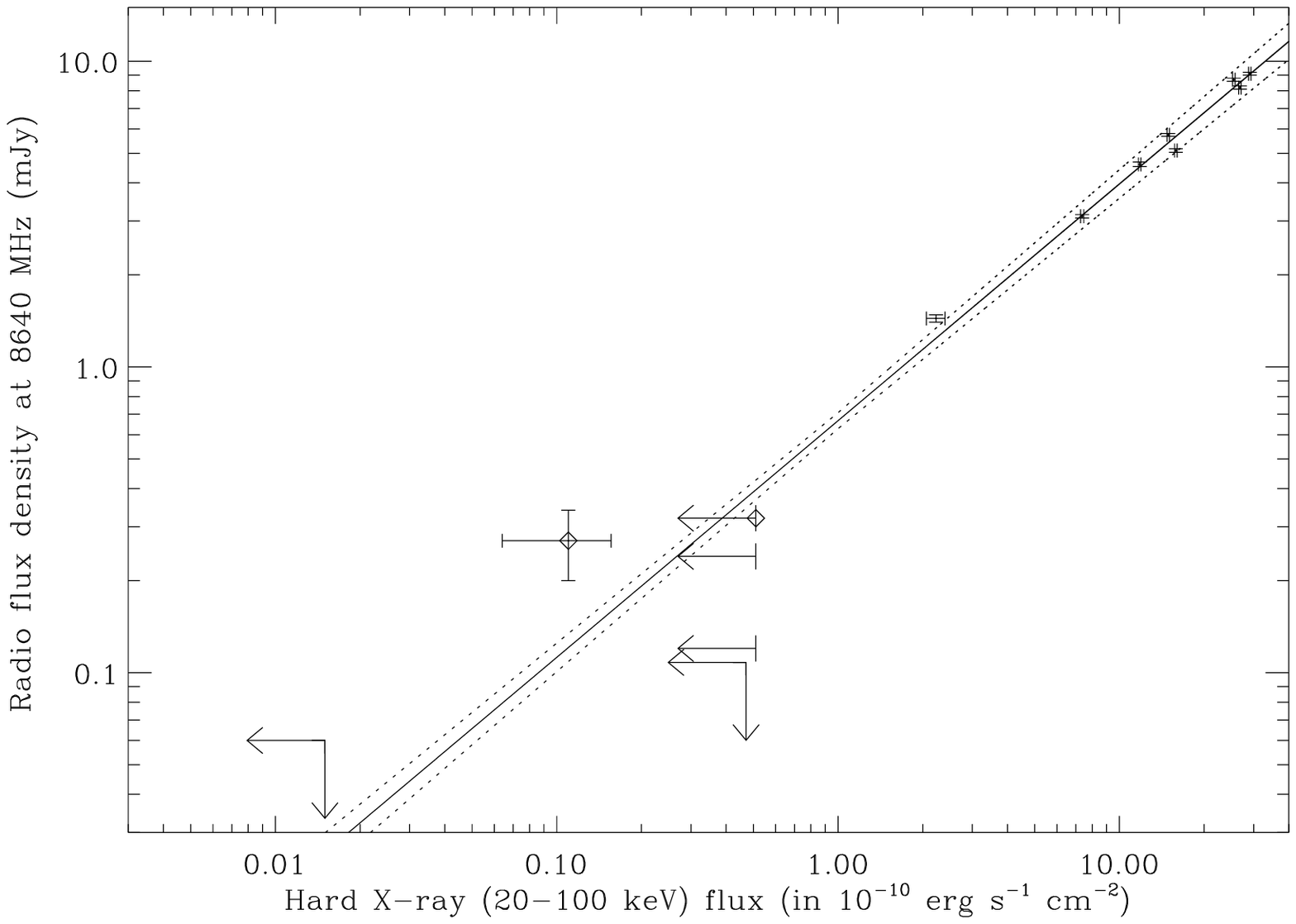}
\caption{Same as Fig. 1, but for the X-ray flux in the 20-100 keV energy band.}
\end{figure*}

\begin{figure*}[B] \centering \includegraphics[width=14cm]{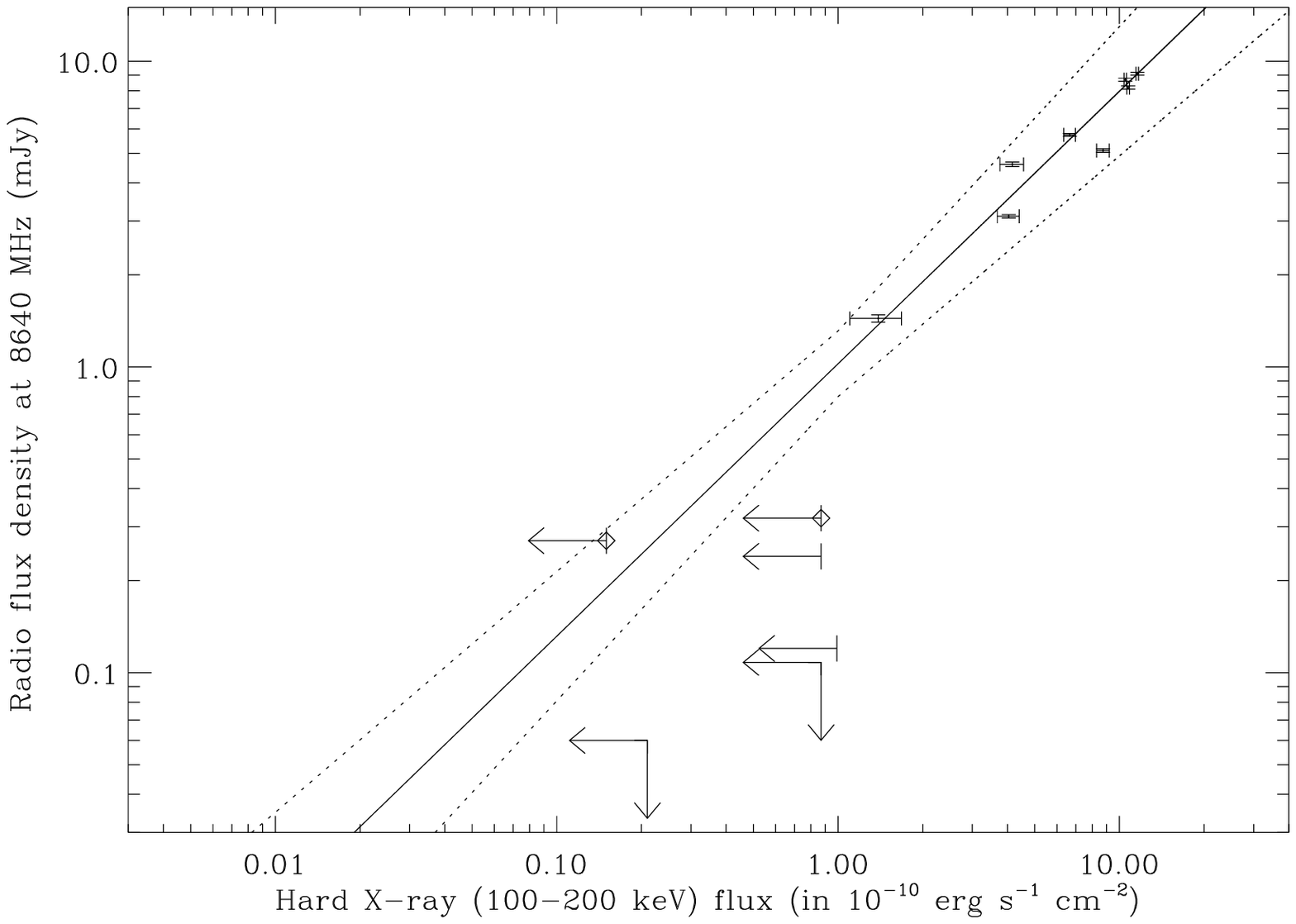}
\caption{Same as Fig. 1, but for the X-ray flux in the 100-200 keV energy band.}
\end{figure*}

\end{document}